\documentclass[a4paper,11pt]{article}
\usepackage{pos}
\usepackage{subfigure}   
\usepackage{subcaption}
\usepackage{graphicx}
\usepackage{sidecap}
\usepackage{slashed}
\usepackage{eucal}
\usepackage{booktabs} 
\usepackage{siunitx} 
\DeclareSIUnit\barn{b} 

\usepackage{xcolor}
\usepackage[normalem]{ulem}

\definecolor{ctcolorblack}{gray}{0}
\definecolor{ctcolorgray}{gray}{.5}
\definecolor{ctcolorgraylight}{gray}{.8}
\definecolor{ctcolorgraylighter}{gray}{.95}

\usepackage{titlesec}

\titlespacing*{\section}
  {0pt}   
  {8pt}   
  {5pt}    

\titlespacing*{\subsection}
  {0pt}
  {8pt}
  {4pt}

\usepackage{listings}

\title{Bound-state production in MadGraph5\_aMC@NLO}

\author*[a]{Alice Colpani Serri}
\emailAdd{alice.colpani\_serri.dokt@pw.edu.pl}
\author[b,c]{Chris A. Flett}
\emailAdd{christopher.flett@uclouvain.be}
\author[b]{Jean-Philippe Lansberg}
\emailAdd{Jean-Philippe.Lansberg@in2p3.fr}
\author[c]{Olivier Mattelaer}
\emailAdd{olivier.mattelaer@uclouvain.be}
\author[d]{Hua-Sheng Shao}
\emailAdd{huasheng.shao@lpthe.jussieu.fr}
\author[d]{Lukas Simon}
\emailAdd{lsimon@lpthe.jussieu.fr}

\affiliation[a]{Faculty of Physics, Warsaw University of Technology, plac Politechniki 1, 00-661, Warszawa, Poland}
\affiliation[b]{Universit\'e Paris-Saclay, CNRS, IJCLab, 91405 Orsay, France}
\affiliation[c]{Centre for Cosmology, Particle Physics and Phenomenology (CP3), Universit\'e Catholique de Louvain, Chemin du Cyclotron, Louvain-la-Neuve,
B-1348, Belgium}
\affiliation[d]{Laboratoire de Physique Th\'eorique et Hautes Energies (LPTHE), UMR 7589, Sorbonne Universit\'e et CNRS, 4 place Jussieu, 75252 Paris Cedex 05, France}

\abstract{
I present the first implementation in MadGraph5\_aMC@NLO for bound-state production, including quarkonium, leptonium, and $B_c$ mesons within Non-Relativistic Quantum Chromo-Dynamics (NRQCD) and Non-Relativistic Quantum Electro-Dynamics (NRQED). In these proceedings, I focus on the extension of MadGraph5\_aMC@NLO to quarkonium states and describe the capabilities of the new framework for inclusive and associated S-wave quarkonium production in a wide variety of experimental environments at Leading Order (LO).
These developments provide the community with a unified, automated, and efficient framework for bound-state production in QED and QCD, and establish a solid foundation for our future Next-to-Leading-Order (NLO) extension. }

\FullConference{33rd International Workshop on Deep Inelastic Scattering and Related Subjects (DIS2026)\\
 4-8 May 2026\\
Bologna, Italy\\}

\AtBeginDocument{%
  \setlength{\abovedisplayskip}{1pt}
  \setlength{\belowdisplayskip}{1pt}
  \setlength{\abovedisplayshortskip}{1pt}
  \setlength{\belowdisplayshortskip}{1pt}
}

\begin{document}
\maketitle

\vspace*{-1cm}
\section{Introduction}

\noindent
\texttt{MadGraph5\_aMC@NLO} (\texttt{MG5}) \cite{Alwall:2014hca} 
is a Monte-Carlo~(MC) event-generator which computes (differential) cross sections and generates particle events. It provides a complete set of tools to generate Feynman diagrams, compute their amplitudes and finally obtain cross sections in a fully-automated way.
It provides a flexible framework to support a large number of particle physics models within the Standard Model (SM) and beyond (BSM). The cross section of an inclusive process $h_A+ h_B \rightarrow k + X$, with $h_{A,B}$ two hadrons in the initial state and  $k$ the final-state elementary particle(s), is computed in \texttt{MG5} through the collinear-factorisation formalism as

\vspace{-0.3cm}
\begin{equation}\label{eq:xsec}
    {\rm d}\sigma(h_A + h_B \rightarrow k + X) = \sum_{a,b} \int {\rm d}x_a {\rm d}x_b 
    f_{a/{h_A}}(x_a) f_{b/{h_B}}(x_b) {\rm d}\hat{\sigma}(ab \rightarrow k + X) (x_a,x_b,\mu_F,\mu_R) \,,
\end{equation}
 where $a$ ($b$) is the parton from hadron $h_A$ ($h_B$) and $x_{a}$ ($x_b$) is the momentum fraction of the parent hadron carried by $a$ ($b$). 
 $f_{a/{h_A}}(x_a)$ and $f_{b/{h_B}}(x_b)$ are the Parton Distribution Functions (PDFs) and ${\rm d}\hat{\sigma}(ab \rightarrow k + X)$ is the partonic cross section, which depends on parton momenta, factorisation $(\mu_F)$ and renormalisation $(\mu_R)$ scales. 
Our work has been devoted to the extension of this framework to bound-state production processes. To do so, we follow the NRQCD formalism \cite{Bodwin:1994jh}, within which the cross section of the process $h_A+ h_B \rightarrow \mathcal{B} + X$, with $\mathcal{B}$ a non-relativistic bound state, can be computed further factorising Eq.~\eqref{eq:xsec} as

\vspace{-0.1cm}
\begin{equation}\label{eq:xsecQ}
    {\rm d}\sigma( h_A+h_B \rightarrow \mathcal B + X) = \sum_{a,b,n} \int {\rm d}x_a {\rm d}x_b f_{a/{h_A}}(x_a) f_{b/{h_B}}(x_b) {\rm d}\hat{\sigma}(ab \rightarrow (C_1 C_2)[n] + X_p) \langle \mathcal O_{n}^{\mathcal B} \rangle \,.
\end{equation}

\vspace{-0.1cm}
\noindent
Here, ${\rm d}\hat{\sigma}$ is the partonic cross section to produce the constituents $C_1$ and $C_2$ in a quantum state $n$, which then evolves into the physical bound state $\mathcal B$ with a probability encoded in the long-distance matrix element (LDME) $\langle \mathcal O_n^{\mathcal B} \rangle$. Final bound states made of two constituent particles can be quarkonia, leptonia or $B_c$ mesons.
Quarkonium states are composed of a heavy quark and its corresponding antiquark, $(C_1 C_2) = (Q \bar{Q}^\prime)$ and $\mathcal B = \mathcal Q$, and are the focus of these proceedings. For completeness, leptonium states are composed of a pair of massive, oppositely-charged leptons, $(C_1 C_2) = (\ell^- \ell^{\prime +})$ and $\mathcal B = \mathcal L$. A complete discussion of the \texttt{MG5} extension including leptonium production and $B_c$ mesons is available in~\cite{serri2025automatedeventgenerationswave}, whereas here we consider the case $\mathcal B = \mathcal Q$ only.

\section{Quarkonium production in NRQCD}

\noindent
Quarkonium states are among the simplest bound states in QCD, and serve as great tools for QCD studies from both a theoretical and an experimental point of view~\cite{Lansberg:2019adr}.
Within NRQCD, the cross section for a quarkonium production process is defined as Eq.~\eqref{eq:xsecQ}, where $ C_1C_2[n] = Q \Bar{Q}[n]$ in the partonic cross section is the open heavy quark and antiquark pair in the state $n$ defined as

\vspace{-0.5cm}
\begin{equation}\label{eq:spec}
    n ={}^{2S+1}L^{[C]}_{J} \,,
\end{equation}
using spectroscopic notation.
The quantum numbers are $S$ for the spin, $L$ for the orbital angular momentum, $J$ for the total angular momentum and $C$ for the colour.
The partonic cross section for such a process can be calculated through the relation

\vspace{-0.5cm}
\begin{equation}
    {\rm d}\hat{\sigma}(ab \rightarrow (Q \Bar{Q})[n] + X_p) \sim |\mathcal{A}_{\{ [C], S, L, J \}}|^2 \,,
\end{equation}

\noindent
where $\mathcal{A}_{\{ [C], S, L, J \}}$ is obtained by projecting the amplitude $\mathcal{A}$ of the partonic process with

\vspace{-0.3cm}
\begin{equation}
    \mathcal{A}_{\{ [C], S, L, J \}} = \sum_{\lambda_S \lambda_L}\mathbb{P}_J \Big[ \mathbb{P}_L \sum_{\lambda_Q \lambda_{\Bar{Q}}}\mathbb{P}_S\mathbb{P}_{[C]} \mathcal{A}  \Big]_{q=0} \,.
\end{equation}
Here, $\mathbb{P}_{C}$ projects the heavy-quark pair onto a colour-singlet ($\mathbb{P}_{C=1}= \delta_{c_1 c_2}/\sqrt{N_c}$~\footnote{$N_c=3$ in QCD.}) or colour-octet ($\mathbb{P}_{C=8} = \sqrt{2}\,t_{c_1 c_2}^{c_{12}}$) Fock state, with $c_1$ and $c_2$ the color of each heavy quark, and $t_{c_2 c_1}^{c_{12}}$ the Gell-Mann matrix element. $\mathbb{P}_{S}$ is the spin projector, defined as

\vspace{-0.2cm}
\begin{equation}
    \mathbb{P}_S = \frac{\bar{v}_{\lambda_{\Bar{Q}}}(k_{\Bar{Q}})\Gamma_{\!S}\,u_{\lambda_{Q}}(k_{Q})}{2\sqrt{2m_{Q}m_{\Bar{Q}}}}\,,
\end{equation} 
with $\Gamma_{\!S=0}=\gamma_5$ for a spin-singlet state and $\Gamma_{\!S=1}=\slashed{\varepsilon}^{*}_{\lambda_s}(K)$ for a spin-triplet state, with $\varepsilon^{*}_{\lambda_s}(K)$ the polarisation vector of the heavy-quark pair with total four-momentum $K = k_Q + k_{\Bar{Q}}$ and spin-related helicity $\lambda_s = \pm1,0$. In the limit $q \rightarrow 0$, with $q$ the relative momentum of the heavy-quark pair, the orbital angular momentum projector is 

\vspace{-0.6cm}
\begin{equation}
    \mathbb{P}_L = \bigg( \varepsilon_{\lambda_l}^{\mu,*}(K) \frac{{\rm d}}{{\rm d}q^\mu}\bigg)^L \,,
\end{equation}
where $\varepsilon_{\lambda_l}^{\mu,*}(K)$ is the polarisation vector and $\lambda_l$ its polarisation components. 
Finally, the total angular momentum projector is defined as the Clebsch–Gordan coefficient

\vspace{-0.4cm}
\begin{equation}
    \mathbb{P}_J = \langle J, \lambda_j | L, \lambda_l; S, \lambda_s \rangle \,,
\end{equation}
where $\lambda_j = -J, -J+1, \dots, J-1, J$.
For S-wave states ($L=0$), the projection procedure is limited to $\mathbb{P}_C$ and $\mathbb{P}_S$ and is the focus of these proceedings (for more details, we refer to \cite{serri2025automatedeventgenerationswave}). The implementation of P-wave states ($L=1$) has recently been carried out using a dual-number approach, which opens up another layer of phenomenological studies \cite{Maxia:2026ved}.

\section{Implementation}

\noindent
In \texttt{MG5}, bound states are now implemented through the new \texttt{UFO} model called \texttt{sm\_onia} and treated similarly to the \texttt{multiparticles} class: instead of an ensemble of elementary particles, bound states are defined as a superposition of all their Fock states. For example, the $J/\psi$ is defined as\footnote{The corresponding P-wave states are displayed as well, but are not considered in what follows.}

\begin{lstlisting}[basicstyle=\scriptsize\ttfamily, xrightmargin=-0.3em]
# Syntax: label = Fock states (separated by spaces)
Jpsi = Jpsi(1|3S11) Jpsi(1|1S08) Jpsi(1|3S18) Jpsi(1|3P08) Jpsi(1|3P18) Jpsi(1|3P28)
\end{lstlisting}

\noindent
where {\tt Jpsi} is the physical bound state defined via its Fock states, which contain the name of the particle (here {\tt Jpsi}) associated with the corresponding quantum state in round brackets. On the left of the vertical bar is the principal quantum number, followed on the right by the state $n$ in spectroscopic notation, following Eq.~\eqref{eq:spec}. Explicitly, from left to right, $2S+1$ ({\tt 1} or {\tt 3}),
the letter \texttt{S} (for $L=0$), $J$ (here {\tt 0} or {\tt 1}) and lastly the colour $C$ ({\tt 1} or {\tt 8}).
It is possible to list all bound states directly from \texttt{MG5} terminal, with the command \texttt{display boundstates}. The Fock states are defined one by one in \texttt{boundstates.py} through the new class \texttt{Boundstate} containing all information related to a specific configuration. For example, for \texttt{Jpsi(1|3S11)}:

\begin{lstlisting}[basicstyle=\scriptsize\ttfamily]
psi_13s11 = Boundstate(
            pdg_code = 443,
            name = 'Jpsi(1|3S11)',
            particles = ['c','c~'],
            mass = 3.1,
            principal = 1,
            spin = 3,
            orbital = 0,
            J = 1,
            color = 1,
            charge = 0,
            ldme = 1.16,
            texname = 'jpsi13S11')
\end{lstlisting}
Similarly, it is possible to list all Fock states from the \texttt{MG5} terminal, with the command \texttt{display fockstates}.
The LDMEs are defined in the input file \texttt{onia\_card.dat} in \texttt{Block ldme}, where each Fock state is associated with a different LDME value. Following the example for \texttt{Jpsi}:

\begin{lstlisting}[basicstyle=\scriptsize\ttfamily]
Block ldme
   443      1.160000000000000   # LDME for Jpsi(1|3S11)
   9940003  0.009029230000000   # LDME for Jpsi(1|3S18)
   9941003  0.014600000000000   # LDME for Jpsi(1|1S08)
\end{lstlisting}
where the default values for $S$-wave production are shown. The user can change them by modifying the dedicated card when generating the process. With the P-wave update, in \texttt{onia\_card.dat} an additional block containing the mass of the quarkonium has been implemented, which need not be the sum of the quark constituent masses as it was for the S-wave case.  
The usage is explained in more detail in the following section.

\section{Usage instructions}

\noindent
From the \texttt{mg5amcnlo} folder, the command \texttt{./bin/mg5\_aMC} opens the \texttt{MG5} interface.
Let us consider the process $p\,p \rightarrow J/\psi$(3S11). To generate processes with quarkonia in the final state, the new model \texttt{sm\_onia} must be imported. By default the model employs a 4-flavour scheme (4FS), i.e. it considers the charm quark massless, which is fine for production of bottomonia. However, if charmonia are produced (as in the case of $J/\psi$), the flag \texttt{-c\_mass} must be added to the model, to consider a 3FS with a massive charm quark:

{\tt MG5\_aMC> import model sm\_onia-c\_mass

MG5\_aMC> generate p p > Jpsi(1|3S11) 

MG5\_aMC> output folder\_name; launch
}

\noindent
This generates the process and saves it inside \texttt{folder\_name}. After launch, the user will be asked to modify settings and cards in an interactive way, just typing the specific number to select the corresponding option (it is otherwise possible to press enter without any number to bypass these steps). Instead, to generate the process considering the physical bound state, the user can type

{\tt MG5\_aMC> generate p p > Jpsi}

\noindent
which by itself encodes the set of commands (here reported S-wave states only)

{\tt MG5\_aMC> generate p p > Jpsi(1|3S11) 

MG5\_aMC> add process p p > Jpsi(1|1S08) 

MG5\_aMC> add process p p > Jpsi(1|3S18) }

\noindent 
and generate all three subprocesses together. Our implementation handles processes with an arbitrary number of (different) final-state quarkonia, in association with any elementary particles: to generate more complicated processes, simply type the name of the additional particle(s) separated by spaces:  {\tt MG5\_aMC> generate p p > Jpsi Jpsi g}

\section{Benchmarking and results of the S-wave extension}

\noindent
The implementation has been first tested against \texttt{Helac-Onia} \cite{Shao_2013, shao2015helaconia20upgradedmatrixelement}. The benchmarking has been performed first checking the matrix-element squared from both frameworks imposing the same setting conditions, using the \texttt{standalone mode} in \texttt{MG5} (this is done by specifying \texttt{output standalone}, which carries out the computation directly on the terminal up to the matrix-element squared, without performing the whole cross-section calculation). After that, a benchmarking of the cross sections between the two frameworks has been performed, fixing identical input data, such as the kinematic set-up and LDME values, obtaining an excellent agreement. The exhaustive list of tests performed is given in the supplementary files alongside the arXiv submission of our \cite{serri2025automatedeventgenerationswave}. The results of a variety of processes are given below. It is important to keep in mind that the hierarchy of the processes shown highly depends on the LDME values set, and the cuts applied are not specific to any experimental setup. The flexibility of the code allows the user to assign a different LDME to each Fock-state contribution on the fly in an intuitive and straightforward way.
As a first application, we consider inclusive single-quarkonium production in $pp$ collisions at 13 TeV.\footnote{We take the bottom mass $m_b=4.7$~GeV, charm mass (while in the 3FS) $m_c=1.55$~GeV and \texttt{PDF4LHC21\_40}~\cite{PDF4LHCWorkingGroup:2022cjn} NNLO PDF set for all of our computations.}
A sample of the results is given in the following table, with the statistical uncertainty of the MC in round brackets.

\begin{table}[h]
\centering
\small
\begin{tabular}{lc@{\qquad}lc}
\toprule
\textbf{Process} & $\sigma$ & \textbf{Process} & $\sigma$ \\
\midrule
$pp\to\eta_c+X$  & $2.9366(5)\,\mu\mathrm{b}$ &
$pp\to\eta_b+X$  & $5.4935(7)\,\mu\mathrm{b}$ \\
$pp\to J/\psi+X$ & $536.14(6)\,\mathrm{nb}$ &
$pp\to\Upsilon+X$ & $6.0655(4)\,\mathrm{nb}$ \\
\bottomrule
\end{tabular}
\end{table}

\noindent
As second case of study, we consider associated quarkonium production, with an additional elementary particle in the final state, such as $W^{\pm}$ and $Z$ bosons, photons ($\gamma$), but also jets $j$.
A few results are given in the following table, where the cuts used are: $p_{T,j} > 10\,\mathrm{GeV}$ and $|\eta_{j}| < 5$ for the jet and $p_{T,\gamma} > 2\,\mathrm{GeV}$ and $|\eta_{\gamma}| < 2.5$ for the photon.

\begin{table}[h!]
\centering
\small
\begin{tabular}[t]{lc|lc}
\toprule
\textbf{Process}& $\sigma$ & \textbf{Process}& $\sigma$  \\
\midrule
$pp \to J/\psi+j+X$ & $329.8(2)\,\mathrm{nb}$ & $pp \to \Upsilon+j+X$ & $19.85(1)\,\mathrm{nb}$ \\
$pp \to \eta_c+\gamma+X$ & $789.3(4)\,\mathrm{pb}$ & $pp \to \eta_b+\gamma+X$ & $2.257(1)\,\mathrm{pb}$  \\
\bottomrule
\end{tabular}
\end{table}

\noindent
The third case study considers a selection of processes with two final-state quarkonia, either in the same state or mixed. The results are given in the following table.
\begin{table}[h!]
\centering
\small
\begin{tabular}[t]{lc|lc}
\toprule
\textbf{Process}& $\sigma$ & \textbf{Process}& $\sigma$  \\
\midrule
$pp \to \eta_c+\eta_c+X$ & $ 35.81(1)\,\mathrm{nb}$ & $pp \to \eta_b+\eta_b+X$ & $ 75.64(3)\,\mathrm{pb}$ \\
$pp \to \eta_c+J/\psi+X$ & $ 7.233(3)\,\mathrm{nb}$ & $pp \to \eta_b+\Upsilon+X$ & $ 1.9244(6)\,\mathrm{pb}$ \\
$pp \to J/\psi+J/\psi+X$ & $ 10.756(3)\,\mathrm{nb}$ & $pp \to \Upsilon+\Upsilon+X$ & $ 44.63(1)\,\mathrm{pb}$ \\
\bottomrule
\end{tabular}
\end{table}

\noindent
A more complicated case is given by triple-quarkonium production. We consider triple-$J/\psi$ as a demonstration of the robustness of our extension in handling larger multiplicity final states:
\begin{equation*}
    \sigma( p p \rightarrow J/\psi + J/\psi + J/\psi + X) = 1.038(3)~\text{pb}
\end{equation*}

\begin{figure}[h]
\centering
\begin{minipage}{.5\textwidth}
  \centering
  \includegraphics[width=.89\linewidth]{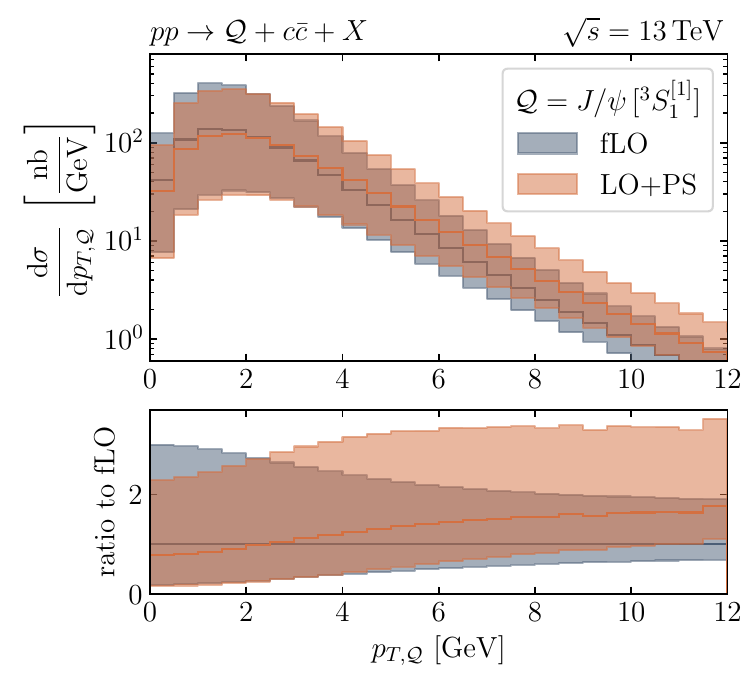}
\end{minipage}%
\begin{minipage}{.5\textwidth}
  \centering
  \includegraphics[width=.89\linewidth]{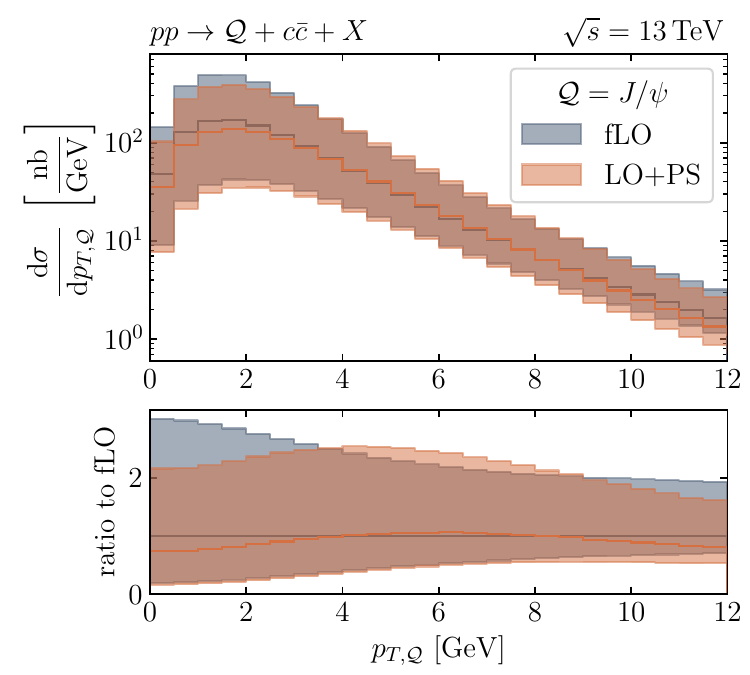}
\end{minipage}
\caption{Differential cross section as a function of $p_{T,\mathcal Q}$ for $pp \rightarrow J/\psi + c \bar{c} + X$ at 13 TeV. Results display production of the colour-singlet $J/\psi [{}^3S_1^{[1]}]$ Fock-state contribution (left panel) and the sum over all Fock-state contributions (right panel) with fixed LO (blue) and LO with PS (orange) contributions.}
\label{fig:PS}
\end{figure}

\noindent
Among \texttt{MG5} convenient features is its compatibility and integration with other softwares. As a demonstration, we study the process $pp \rightarrow J/\psi + c \Bar{c} + X$ with our extension in \texttt{MG5} and interface it with \texttt{Pythia8}~\cite{bierlich2022comprehensiveguidephysicsusage} to parton shower (PS) events.
The LHE files generated by \texttt{MG5} can be directly fed to \texttt{Pythia8} to perform the analysis, where we keep the settings as default. In Fig.~\ref{fig:PS}, we show the plots of the differential cross section as a function of the $J/\psi$ transverse momentum $p_{T,J/\psi}$ at fixed-order and with the parton shower.

\vspace{-1cm}
\noindent
As a final example, we consider (vector) quarkonium production in association with a Higgs boson $H$. For this study, a different model is considered: we employ Higgs Effective Field Theory (HEFT) within a new dedicated \texttt{UFO} model, called \texttt{heft\_onia}. The cross section results are
\begin{equation}
    \begin{aligned}
        \sigma(pp\to gg \to J/\psi+H) &= 1.53^{+0.40}_{-0.29}\,\mathrm{fb}\,,\\
        \sigma(pp\to gg \to \Upsilon+H) &= 91^{+23}_{-17}\,\mathrm{ab}\,.
    \end{aligned}
\end{equation}

\begin{figure}[h]
\centering
\begin{minipage}{.5\textwidth}
  \centering
  \includegraphics[width=.89\linewidth]{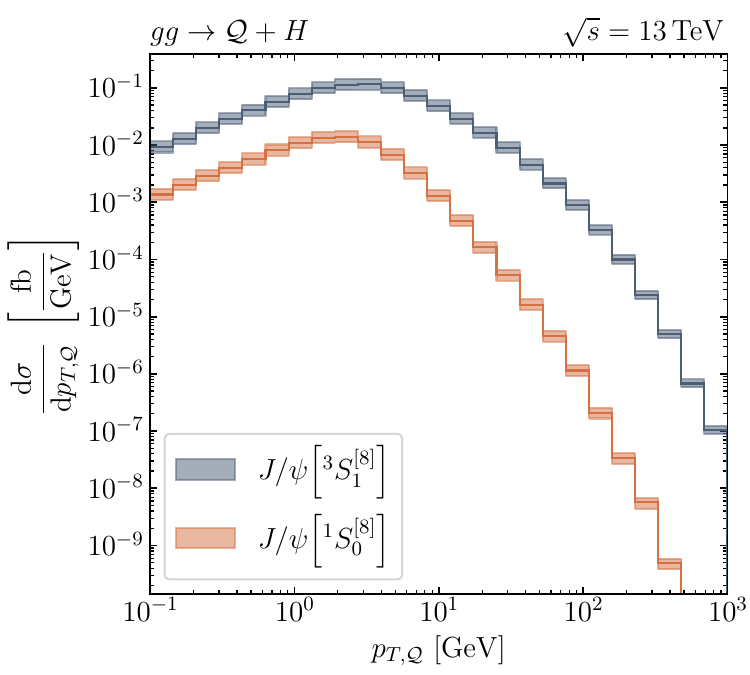}
\end{minipage}%
\begin{minipage}{.5\textwidth}
  \centering
  \includegraphics[width=.89\linewidth]{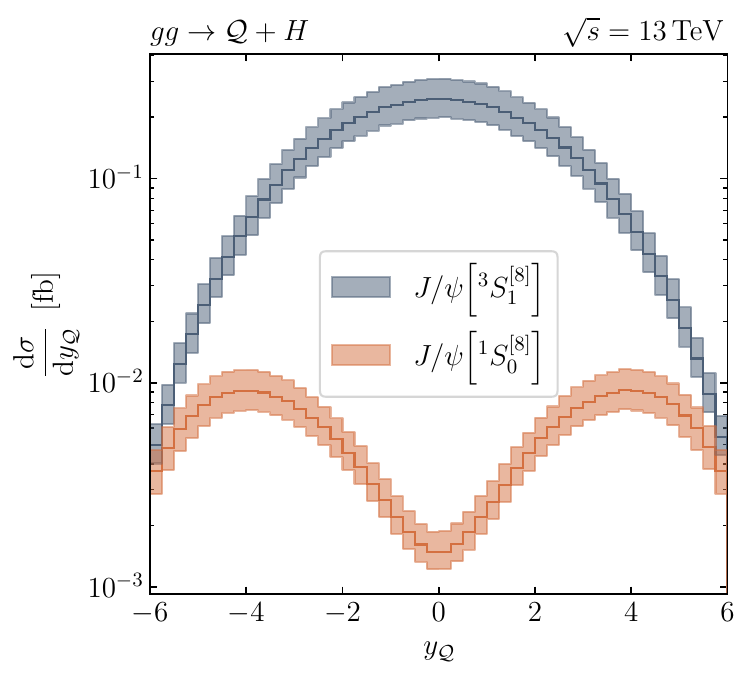}
\end{minipage}
\caption{Transverse momentum (left) and rapidity (right) differential cross-section distributions for $J/\psi + H$ production in $pp$ collisions at $\sqrt{s}=13\,\mathrm{TeV}$.  The contributions from the $^1S_0^{[8]}$ (orange) and $^3S_1^{[8]}$ (blue) Fock states are displayed with error bands from a 7-point scale variation.}
\label{fig:H}
\end{figure}

\noindent
In Fig.~\ref{fig:H}, we consider the $J/\psi$ case and show the plots of the differential cross section as a function of $J/\psi$ transverse momentum $p_{T,J/\psi}$ and rapidity $y_{J/\psi}$.

\section{Conclusions and outlook}

\noindent
\texttt{MadGraph5\_aMC@NLO} (\texttt{MG5}) has been extended to support S-wave \cite{serri2025automatedeventgenerationswave} and P-wave \cite{Maxia:2026ved} bound states at LO in the NRQCD formalism. The implementation enables automated computation of cross sections for a wide variety of processes. These proceedings focused on quarkonium states, however leptonium states and $B_c$ mesons are also supported. A benchmark of the implementation has been done against \texttt{Helac-Onia}, at the level of the matrix-element squared to the full cross-section. The sample of results reported covers a wide range of particle processes, showing the potential and flexibility of our extension, and its compatibility with other \texttt{MG5} features, like its interface with \texttt{Pythia8} and user-defined \texttt{UFO} models. An extension to NLO accuracy is planned, which will open the way to a broad range of phenomenological applications, including, for example, global NRQCD analyses.
The implementation of TMD factorisation, not yet supported by \texttt{MG5}, is under development. At present, all computations are performed within the collinear factorisation framework, and the extension of TMD factorisation is being implemented in the helicity-amplitude formalism.
The (S- and P-wave) implementation will be released with \texttt{MG5} version 3.8.0, and will be publicly available via the standard \texttt{MG5} distribution on launchpad (\url{https://launchpad.net/mg5amcnlo}) and on the NLOAccess EU platform~\cite{Flore:2023dps} (\url{https://nloaccess.in2p3.fr}). 

\section{Acknowledgements}
We would like to thank C.~Flore, R.~Frederix, K.~Lynch, F.~Maltoni, M.~Mangano, M.~Nefedov, and H.-F.~Zhang for useful discussions. L.S. thanks the Centre for Cosmology, Particle Physics and Phenomenology (CP3) at Universit\'e Catholique de Louvain for hospitality, where part of this work was carried out.

\noindent
C.F. acknowledges support from the Marie Skłodowska-Curie Action (``AutomOnium''), funded by the European Union under grant agreement No. 101204057.

\noindent
C.F., J.-P.L., and H.-S.S. are supported by the Agence Nationale de la Recherche (ANR) via the grant ANR-20-CE31-0015 (``PrecisOnium'').

\noindent
The work of C.F. and J.-P.L. is supported by the IDEX Paris-Saclay ``Investissements d'Avenir'' (ANR-11-IDEX-0003-01) through the GLUODYNAMICS project funded by the ``P2IO LabEx (ANR-10-LABX-0038)'', the French CNRS via the IN2P3 projects ``GLUE@NLO” and ``QCDFactorisation@NLO'' as well as via the COPIN-IN2P3 project \#12-147 ``kT factorisation and quarkonium production in the LHC era''.

\noindent
O.M. and the \texttt{MadGraph5\_aMC@NLO} project are supported by FRS-FNRS (Belgian National Scientific Research Fund) IISN projects 4.4503.16 (MaxLHC) and DR-Weave grant FNRS-DFG num\'ero T019324F (40020485). 

\noindent
The work of H.-S.S. and L.S. is supported by the ERC grant 101041109 (``BOSON''). Views and opinions expressed are however those of the authors only and do not necessarily reflect those of the European Union or the European Research Council Executive Agency. Neither the European Union nor the granting authority can be held responsible for them.


\begin{thebibliography}{99}

\bibitem{Alwall:2014hca}
J. Alwall, R. Frederix, S. Frixione, V. Hirschi, F. Maltoni, O. Mattelaer,
H.-S. Shao, T. Stelzer, P. Torrielli and M. Zaro,
\emph{The automated computation of tree-level and next-to-leading order differential cross sections, and their matching to parton shower simulations},
\href{https://doi.org/10.1007/JHEP07(2014)079}
{\emph{JHEP} \textbf{07} (2014) 079}
[{\tt arXiv:1405.0301 [hep-ph]}].

\bibitem{Bodwin:1994jh}
G.T. Bodwin, E. Braaten and G.P. Lepage,
\emph{Rigorous QCD analysis of inclusive annihilation and production of heavy quarkonium},
\href{https://doi.org/10.1103/PhysRevD.51.1125}
{\emph{Phys. Rev. D} \textbf{51} (1995) 1125}
[{\tt hep-ph/9407339}].
[Erratum:
\href{https://doi.org/10.1103/PhysRevD.55.5853}
{\emph{Phys. Rev. D} \textbf{55} (1997) 5853}.]

\bibitem{serri2025automatedeventgenerationswave}
A.C. Serri, C.A. Flett, J.-P. Lansberg, O. Mattelaer, H.-S. Shao and L. Simon,
\emph{Automated event generation for S-wave quarkonium and leptonium production in NRQCD and NRQED},
[{\tt arXiv:2510.26773 [hep-ph]}].

\bibitem{Lansberg:2019adr}
J.-P. Lansberg,
\emph{New Observables in Inclusive Production of Quarkonia},
\href{https://doi.org/10.1016/j.physrep.2020.08.007}
{\emph{Phys. Rept.} \textbf{889} (2020) 1--106}
[{\tt arXiv:1903.09185 [hep-ph]}].

\bibitem{Maxia:2026ved}
L. Maxia, H.-S. Shao and L. Simon,
\emph{Automated NRQCD and NRQED simulations of quarkonium and leptonium production with P-wave states and physical-mass effects},
[{\tt arXiv:2607.26739 [hep-ph]}].

\bibitem{Shao_2013}
H.-S. Shao,
\emph{HELAC-Onia: An automatic matrix element generator for heavy quarkonium physics},
\href{https://doi.org/10.1016/j.cpc.2013.05.023}
{\emph{Comput. Phys. Commun.} \textbf{184} (2013) 2562--2570}.

\bibitem{shao2015helaconia20upgradedmatrixelement}
H.-S. Shao,
\emph{HELAC-Onia 2.0: an upgraded matrix-element and event generator for heavy quarkonium physics},
[{\tt arXiv:1507.03435 [hep-ph]}].

\bibitem{PDF4LHCWorkingGroup:2022cjn}
R.D. Ball \emph{et al.} [PDF4LHC Working Group],
\emph{The PDF4LHC21 combination of global PDF fits for the LHC Run III},
\href{https://doi.org/10.1088/1361-6471/ac7216}
{\emph{J. Phys. G} \textbf{49} (2022) 080501}
[{\tt arXiv:2203.05506 [hep-ph]}].

\bibitem{bierlich2022comprehensiveguidephysicsusage}
C. Bierlich, S. Chakraborty, N. Desai, L. Gellersen, I. Helenius, P. Ilten,
L. Lönnblad, S. Mrenna, S. Prestel, C.T. Preuss, T. Sjöstrand, P. Skands,
M. Utheim and R. Verheyen,
\emph{A comprehensive guide to the physics and usage of PYTHIA 8.3},
[{\tt arXiv:2203.11601 [hep-ph]}].

\bibitem{Flore:2023dps}
C. Flore,
\emph{NLOAccess: automated online computations for collider physics},
\href{https://doi.org/10.1140/epja/s10050-023-00972-2}
{\emph{Eur. Phys. J. A} \textbf{59} (2023) 46}
[{\tt arXiv:2301.09167 [hep-ph]}].


\end{thebibliography}
\end{document}